\documentclass[a4paper,11pt]{article}
\pdfoutput=1

\usepackage{jheppub}

\usepackage{xspace}
\usepackage{bm}

\usepackage{color}
\usepackage{float}
\usepackage{dsfont}
\usepackage{graphicx}

\usepackage{epstopdf}
\usepackage{slashed}
\usepackage{array}  
\usepackage[absolute]{textpos}
\usepackage{booktabs}

\input{commands}
\newcommand\blockoftext{{
There is some discussion in the literature regarding how the incorporation of
background components comprising a mean with some uncertainty affects the frequentist coverage
properties of $p$-value limits \cite{2003sppp.conf..261C}. In particular, when
one is considering a background that is constrained e.g.~from an MC sample, the
acceptance region for the hypothesis test in the full Neyman construction will
vary according to the value assumed by the nuisance parameter(s) controlling the
strength of the background.  In an approximated scheme, such as the profiling
method used in the computation of \clsb and \cls, the coverage can hence deviate
from that nominally expected; potentially significantly if the background
overestimates the data.  Since both Methods A and C feed information into the
likelihood in a similar way (and have the shortcoming that the likelihood used
in the limit-setting procedure is not the likelihood for \emph{all} the data),
we should not be surprised if one or both methods under- or over-cover. It is hoped,
however, that by virtue of the \MLE\ fake rate being more `sensible' than that
from the matrix method, any deviations in coverage from that nominally expected
would be less extreme in \methodC than with \methodA. \methodB\ should have the
most accurate coverage, although it still might not be exactly correct due to
the use of profiling.  These expectations will be confirmed in the results which follow.}}

\begin{document}   
\title{
  Improving estimates of the number of `fake'
  leptons and other mis-reconstructed objects in hadron collider events: BoB's your UNCLE\footnote{
      `BoB Method' is short for
    `{\bf B}est bits {\bf o}f {\bf B}oth existing fake rate estimation {\bf Method}s'.  UNCLE stands for ``{\bf Un}-biased {\bf C}onfidence {\bf L}imit {\bf E}valuator''. Both are somewhat artificial acronyms designed to make the name of Method C in Table~\ref{table:methodComparison} memorable.}}
\date{\today}

\author{Thomas P. S. Gillam}
\author{and Christopher G. Lester}
\affiliation{Cavendish Laboratory,\\ Dept of Physics, JJ Thomson Avenue,
Cambridge, CB3 0HE, United Kingdom}

\emailAdd{gillam@hep.phy.cam.ac.uk}
\emailAdd{lester@hep.phy.cam.ac.uk}

\abstract{%
We consider current and alternative approaches to setting limits on new physics signals
having backgrounds from misidentified objects; for example jets misidentified as leptons, \bjets\ or photons. Many ATLAS and CMS
analyses have used a heuristic ``matrix method'' for estimating the background
contribution from such sources.  We demonstrate that the matrix method suffers from
statistical shortcomings that can adversely affect its ability to set robust
limits. A rigorous alternative method is discussed, and is seen to produce fake rate estimates and 
limits with better qualities,
but is found to be too 
costly to use.  Having investigated the nature of the approximations used to derive the matrix method, we propose a third strategy that is seen to marry the speed of the matrix method to the performance and physicality of the more rigorous approach.
}

\arxivnumber{1407.5624}

\dedicated{CAV-HEP-14/08}


\maketitle


\section{Introduction}

Many precision measurements and searches for new physics employ signal regions
which have a significant source of background coming from `fake' objects. A
typical concrete example is that of leptons, which can be faked by a
mis-reconstructed jet.  Alternatively one can consider jets faking \bjets, for
which the matrix method was used in \cite{ATLAS-CONF-2013-061}, or even faking
photons. In this article the term `lepton' shall be used throughout, but all
statements made are general to other types of object.
Properties whose distributions differ for `fake' and
`real' objects have been used to underpin data-driven methods of
fake rate estimation, one of the most prevalent of which during the first
data-taking run of the LHC has been the so-called `matrix method' \cite{ATLAS-CONF-2010-087} used by ATLAS in 
\cite{atlasMMCitation0,atlasMMCitation1,atlasMMCitation2,atlasMMCitation3,atlasMMCitation4,atlasMMCitation5,atlasMMCitation6,atlasMMCitation7,atlasMMCitation8,atlasMMCitation9,atlasMMCitation10,atlasMMCitation11,atlasMMCitation12,atlasMMCitation13,atlasMMCitation14,atlasMMCitation15,atlasMMCitation16,atlasMMCitation17,atlasMMCitation18,atlasMMCitation19,atlasMMCitation20,atlasMMCitation21,atlasMMCitation22,atlasMMCitation23,atlasMMCitation24,atlasMMCitation25,atlasMMCitation26,atlasMMCitation27,atlasMMCitation28},
and by CMS in
\cite{cmsMMCitation0,cmsMMCitation1,cmsMMCitation2} based, apparently, on a description
in \cite{Chatrchyan:2012bra}.

The matrix method is the first of three ways of determining fake rates that are compared in this paper.  We shall sometimes refer to it for short \methodA to facilitate easy comparison with the later Methods B and C. (For quick reference see
\tableref{methodComparison}.)

\methodA makes use of the fact that fake and real leptons tend to
differ in their degree of `isolation'.\footnote{The `isolation' of a charged
object is, in this case, a measure of the amount of activity found close to the
track left by the object in the inner detector.}  Using cuts on isolation (and
to a lesser extent other variables) leptons are categorised as either `tight' or
`loose', the former being largely synonymous with `more isolated' and the latter
with `less isolated'.  In this paper it is shown that the way in which the
resulting matrix method derived background estimates are typically used in
existing SUSY searches can give rise to confidence limits that are unstable (highly variable) indicating they are making non-optimal use of the data. 
More specifically, over the course of many independent experiments one expects to find a distribution of limits from \methodA which has a larger variance (is more widely spread out) than the distribution of limits coming from  Methods B and C discussed later.  In addition, \methodA can produce unphysically negative estimates for fake rates that should be bounded below by zero.

\begin{table}
  \centering
  \renewcommand{\arraystretch}{1.4}
  \begin{tabular}{cccc}
    \toprule
    \textbf{Method} & \textbf{Cost of computation} & \textbf{Limit quality} & \textbf{Other names}                    \\
    \midrule
    \textbf{A}      & low                          & poor                   & `Matrix  Method' \cite{ATLAS-CONF-2010-087}                  \\
    \textbf{B}      & very high                    & very good              & `Likelihood Matrix Method'  \\
    \textbf{C}      & quite low                    & good                   &
    `BoB Method'      \\
    \bottomrule
  \end{tabular}
  \caption{
    \label{table:methodComparison}
    An overview of the three methods discussed in this paper, and their relative
    strengths and weaknesses. `Limit quality' refers to whether \clsb limits
    tend to have the correct frequentist coverage properties, and also avoid
    unnecessary over-coverage.
  }
\end{table}

The weaknesses of the matrix method stem from the presence of heuristic,
non-mathematical steps in its derivation.  Heuristic methods are often useful,
but can be problematic if their underlying assumptions are sometimes invalid; we
show this is the case with the matrix method.
Methods lacking these deficiencies
are in principle trivial to construct. In \methodB we describe an example of such a method
in which a single likelihood is used to perform both the background estimation
and the limit setting. Whilst this can be considered the
optimal approach, it is shown to be computationally expensive in cases where
objects are divided into many categories.\footnote{Many categories would be
required if reconstruction efficiencies varied as a function of, say, detector
rapidity or lepton transverse momentum.}    In order to marry the best of Methods A and B, we then propose \methodC.  It is intended that \methodC be usable as a drop-in replacement for \methodA in the contexts in which the latter has previously been used by ATLAS or CMS.  Method C, like \methodA, it is partly heuristic (for speed) and so is justified pragmatically. However, careful choice of the approximations it contains allows it   
to always give physical limits whose distributions very closely resemble the optimal (but prohibitively expensive) \methodB.

Note that fake rate estimates in LHC analyses are likely to find themselves being used as part of a \cls frequentist limit setting procedures, since these are endemic in ATLAS and CMS papers. 
Such usage requires a likelihood for a set of parameters given observed data; in
the case where one counts events in just one region, it typically takes the form
of a Poisson distribution having some mean. Background estimates in an
analysis are then interpreted as an auxiliary measurement which constrains this
mean through additional terms in the likelihood. In this context, the fake
rate ought to be an estimate of the expected number of events from the fake
background process in the signal region, given data collected \emph{outside} of
the signal region. This is not strictly adhered to in the matrix methods, and so is one of the general places in which Methods B and C improve on A.

\section{Overview of fake estimation procedures and limit setting}

Events collected into some signal region are defined in terms of the numbers of
leptons they contain. A cut on some measure of quality,
for example isolation, distinguishes a given lepton as
\emph{loose} (\objloose) or \emph{tight} (\objtight), where $\objloose \cup
\objtight \equiv \objall$ and $\objloose \cap \objtight = \emptyset$. Each
lepton will also be regarded as either \emph{real} (\objreal) or \emph{fake}
(\objfake), depending on whether it is a correctly reconstructed lepton, or for
example a mis-reconstructed heavy flavour jet. According to the precise
selection, a certain number of tight and loose leptons will be required for an
object to make it into a signal region -- those that do are described as
\emph{tight events} (\eventtight), and those that do not but \emph{could} be made to pass the
selection for some permutation of \objtight and \objloose on its constituent
leptons are denoted as loose events (\eventloose), where as before $\eventloose \cup
\eventtight \equiv \eventall$ and $\eventloose \cap \eventtight = \emptyset$. 

A core concept in all methods considered here is that of the real and fake
\emph{efficiencies}, respectively defined to be $\effReal \equiv
P(\objtight|\objreal\objall)$ and $\effFake \equiv P(\objtight|\objfake\objall)$.
For convenience we will also use $\effRealBar \equiv 1-\effReal =
P(\objloose|\objreal\objall)$ and $\effFakeBar \equiv 1-\effFake =
P(\objloose|\objfake\objall)$. Typically these quantities are measured in
additional control regions, and could be subdivided according to kinematic
quantities, such as lepton \pt. In this text such categories will be labelled
$\objcategory{1}, \objcategory{2}, \ldots$, with the efficiencies gaining an
additional subscript e.g. \effRealObj{1}.

For a given event containing \numlep leptons, each lepton is observed to be
either \objloose or \objtight, and will have some category \objcategory{i}. If
there are \numobjcategories possible categories for each lepton, then the number of
measurable \emph{event categories} will be $\numeventcategories = (2\times
\numobjcategories)^\numlep$.\footnote{If the number of leptons can differ between
events, one introduces an appropriate sum over $\numlep$.} Each of these will correspond to an
event that is either \eventloose or \eventtight.

Experimentally, one counts how many events fall into each of the
\numeventcategories sub-regions, yielding the set of integers
$\left\{\numObsInEventCategory{i}\right\}$. For the purpose of the physics
analysis being performed, one might be interested in the total number of
tight events, $\numTight = \sum_{\eventcategory{i}\subset\eventtight}\numObsInEventCategory{i}$.
Usually this is the quantity with which a limit on the cross section of a new
physics model is placed. 

The observed numbers of events are often assumed to be the particular values of
a Poisson distributed random variable. That is, one can have
$\numTight\distPoiss{\rateTight}$; in general the indices on the rate $\nu$
correspond to those on the observation $n$.

\subsection{\methodA: the ``matrix method''}
\label{subsec:matrixMethod}
This section attempts to document \methodA, the matrix method, in more detail
than has previously been done, and in its most general form. As mentioned previously
it is a somewhat heuristic method, but its assumptions shall be interpreted on a
firmer statistical footing in a subsequent section.

\subsubsection{Events with only one lepton}
Consider first a simplified scenario where
each event has exactly one lepton; $n_T$ tight and $n_L$ loose events are
observed. The key relation is then that
\begin{align}
\begin{pmatrix} \expn{T} \\ \expn{L} \end{pmatrix} 
  &= 
  \begin{pmatrix}
    \effReal & \effFake \\
    \effRealBar & \effFakeBar
  \end{pmatrix} 
  \begin{pmatrix} n_R \\ n_F \end{pmatrix}, 
  \label{eqn:basicMatrixMethod}
\end{align}
where $n_R$ and $n_F$ are the number of the observed events which are real and
fake, respectively. In this context, $\expn{L} = \expectation{\numLoose | \numReal,
\numFake}$, and similarly for $\expn{T}$.  The result follows by considering the
real/fake event counts to be random variables following a Poisson distribution,
which are then further divided into tight and loose components according to a
binomial distribution using the probabilities contained in the efficiencies.%

It can be noted that \eqnref{basicMatrixMethod} is similar to a
relation between the means of Poisson distributions
\begin{align}
\begin{pmatrix} \rateTight \\ \rateLoose \end{pmatrix} 
  &= 
  \begin{pmatrix}
    \effReal & \effFake \\
    \effRealBar & \effFakeBar
  \end{pmatrix} 
  \begin{pmatrix} \rateReal \\ \rateFake \end{pmatrix}.
  \label{eqn:basicMatrixMethodForRates}
\end{align}
This is used later when discussing \methodsBandC, but for now we
shall proceed with \eqnref{basicMatrixMethod}. This equation may legitimately be inverted
\begin{align}
  \begin{pmatrix} n_R \\ n_F \end{pmatrix} 
  &= 
  \frac{1}{\effReal - \effFake}
  \begin{pmatrix}
    \effFakeBar & -\effFake \\
    -\effRealBar & \effReal
  \end{pmatrix} 
  \begin{pmatrix} \expn{T} \\ \expn{L} \end{pmatrix},
  \label{eqn:basicMMInversionExact}
\end{align} 
provided that $\effReal \ne \effFake$.  Given the model assumptions that were
made, the steps described hitherto all hold water on mathematical grounds.  In
contrast, the next step that is usually taken to motivate \methodA\ is quite
arbitrary, and is justified largely on grounds that it is effective in
situations with large numbers of events, rather than because it is meaningful in
general.\footnote{We shall see in fact that it is \emph{not} meaningful when the numbers
in question are small.}
This `heuristic' step consists of the removal of the
expectation brackets from the right hand side of \eqnref{basicMMInversionExact}
and the re-interpretation of the terms on its left hand side as a pair of quantities $\hat n_R$ and $\hat n_F$ as follows:
\begin{align}
  \begin{pmatrix} 
\hat n_R \\ \hat n_F 
\end{pmatrix} 
  &=
  \frac{1}{\effReal - \effFake}
  \begin{pmatrix}
    \effFakeBar & -\effFake \\
    -\effRealBar & \effReal
  \end{pmatrix} 
  \begin{pmatrix} n_T \\ n_L \end{pmatrix}. 
  \label{eqn:basicMMAccurateInversion}
\end{align}
What are $\hat n_R$ and $\hat n_F$?  They depend on $n_T$ and $n_L$ and so are
functions of the data, and may be regarded as estimators -- but estimators for
what?  It is shown in \appendixref{origin} that under some additional
assumptions, and for certain values of $n_T$ and $n_L$, they turn out to be
maximum likelihood estimators for $n_R$ and $n_F$ given knowledge of $n_T$ and
$n_L$ ({\it i.e.}~estimators for $\expn{R}\equiv\expectation{\numReal |
\numTight,\numLoose}$ and $\expn{F}\equiv\expectation{\numFake |
\numTight,\numLoose}$).%
\footnote{Note, as shown, that these expectation values
are necessarily conditioned on different things to the expectations values
$\expn{T}$ and $\expn{L}$ seen earlier.}
Nonetheless, and in the absence of anything better, \methodA\ instead uses $\hat n_R$
and $\hat n_F$ as estimators for the unknown and unknowable {\it actual} rates of
real and fake events, \rateReal\ and \rateFake.

Note that these estimators are sometimes pretty bad as equation
\eqref{eqn:basicMMAccurateInversion} allows terms on its left hand side to
become unphysically negative.\footnote{
  For example, consider the case where
  $\numTight=7,\numLoose=1,\effReal=\frac{4}{5}$, and $\effFake=\frac{1}{5}$. One can
  then show that $\hat n_R=9$, and $\hat n_F=-1$.   
}
This happens in real analyses (\textit{e.g.}~\cite{atlasMMCitation2}) creating
problems that need to be solved by \textit{ad-hoc} methods. Both \methodsBandC\
have the benefit of avoiding such problems.

Finally, \methodA\ obtains its desired goal, the definition of an estimator for
the expected number of fake events in the signal region, \ntfhat, motivated by
\eqnref{basicMatrixMethod} with the replacement $n_F\rightarrow\hat n_F$, where
$\hat n_F$ is the estimator obtained above in
\eqref{eqn:basicMMAccurateInversion}.  This results in:
\begin{align}
  \ntfhat &= \frac{\effFake}{\effReal -
    \effFake}\left(\effReal(n_T+n_L)-n_T\right).
  \label{eqn:basicMatrixMethodPrediction}
\end{align}
Again, note the problems with this method that, even if $\effReal>\effFake$ (as
must be the case for a useful definition of \objtight\ and \objloose), 
\eqnref{basicMatrixMethodPrediction} can yield $\ntfhat<0$, an
unphysical result which is symptomatic of the earlier ``sleight of hand''.

This concludes our description of how \methodA\ is
used in single lepton events to calculate a number which is used as if it were
an estimate of the expected rate of fakes in the signal region.  We will now
describe how the same method is extended for use in events with more than one
lepton, which has previously not been documented in detail.

\subsubsection{Events with multiple leptons}
Consider an event with two leptons, where each lepton can be in one of a
number of categories $\{\omega_i\}$.  One may define quantities such as
$n_{tl}$, the number of events with the first\footnote{The definition of
  ``first'' can depend on the analysis. Often it is chosen to be the hardest
according to \pt.} lepton tight and the second loose -- others are defined
similarly. In order to include the possible categories for each lepton, event
counts such as $n_{tt}$
must be further subdivided to take into account all combinations:
\begin{align}
  n_{tt} &= n_{tt}^{\objcategory{1}\objcategory{1}} + n_{tt}^{\objcategory{1}\objcategory{2}} + \cdots \\
         &= \sum_{i,j} n_{tt}^{\objcategory{i}\objcategory{j}}.
  \label{eqn:nttExpansionWithCategories}
\end{align}
In this notation, $n_{tt}^{\objcategory{1}\objcategory{2}}$ indicates the number
of events with two tight leptons, where the first is in category
\objcategory{1}, and the second in $\objcategory{2}$.

The analogous relation to \eqnref{basicMatrixMethod} is then
\begin{align}
  \expected{\tensTL_{\beta_1\beta_2}^{\objcategory{i}\objcategory{j}}} &=
  \fakephi{i}{\beta_1}{\alpha_1}\fakephi{j}{\beta_2}{\alpha_2}\
  \tensRF_{\alpha_1\alpha_2}^{\objcategory{i}\objcategory{j}}, \qquad \text{where}\\
  \fakephi{i}{\{\objtight,\objloose\}}{\{\objreal,\objfake\}} &=
  P(\{\objtight,\objloose\} | \{\objreal,\objfake\}\objcategory{i}\objall),\\
  \phi_\objcategory{i} &= 
    \begin{pmatrix}
      \effRealObj{i} & \effFakeObj{i} \\
      \effRealBarObj{i} & \effFakeBarObj{i}
    \end{pmatrix},
  \label{eqn:twoLeptonMM}
\end{align}
where summation over repeated upper and lower indices is implied where
appropriate.\footnote{In this case a sum should \emph{not} be carried out over
\objcategory{i} or \objcategory{j}, since they appear on the left-hand side.
Despite some notational and behavioural similarities, these objects are not
tensors!} The identifier $n$ has been replaced for clarity with the symbols
$\tensTL$ and $\tensRF$, depending on whether the accompanying indices
pertain to tight/loose-ness, e.g. $\tensTL_{tl} = n_{tl}$, or
real/fake-ness, so $\tensRF_{rf} = n_{rf}$.  Each Greek lower index of $\tensTL$
hence takes values in $\{t,l\}$, while each  Greek lower index of $\tensRF$ takes values in $\{r,f\}$. 
The index on these indices corresponds to the which lepton is being described;
\textit{i.e.} in 
\eqnref{twoLeptonMM}, the value of $\alpha_2$ represents whether the second
lepton is either real or fake. 

The matrix representation for $\phi_\objcategory{i}$ shown in the last line of
\eqnref{twoLeptonMM} is not needed to understand this equation, but is
required when considering the background estimate for events that are both tight
and fake (it is what still identifies this as the ``matrix method'', despite the
new notation).

The estimate for the expected number of \emph{events} that are fake is then
$\tensTLHatFake_{\alpha_1\alpha_2}$, where
\begin{align}
  \tensTLHatFake_{\nu_1\nu_2} = 
  \sum_{i,j}\left(
  \fakephi{i}{\nu_1}{\mu_1} \fakephi{j}{\nu_2}{\mu_2}\ 
  \zeta_{\mu_1\mu_2}^{\ \beta_1\beta_2} \ 
  \fakeinvphi{i}{\beta_1}{\alpha_1} \fakeinvphi{j}{\beta_2}{\alpha_2}
  \tensTL_{\alpha_1\alpha_2}^{\objcategory{i}\objcategory{j}}\right),
  \label{eqn:generalisedMatrixMethod}
\end{align}
where the $\zeta$ object is responsible for defining what is meant by a fake event.
For example, if $rr\equiv\eventreal$ and $\{rf,fr,ff\}\equiv\eventfake$ then one
would choose $\zeta_{12}^{12}=\zeta_{21}^{21}=\zeta_{22}^{22}=1$, and all other
components 0.  There is in fact a redundancy in the indices, in that all
non-zero components have the $i^\text{th}$ lower index the same as the
$i^\text{th}$ upper index. In general therefore, for the case with any number
of leptons
\begin{align}
  \zeta_{\mu_1\mu_2\cdots}^{\ \beta_1\beta_2\cdots} &=
  \delta_{\mu_1}^{\ \beta_1}\delta_{\mu_2}^{\ \beta_2}\cdots 
      h(\beta_1, \beta_2, \ldots),
\end{align}
where $\delta_i^{\ j}$ is the Kronecker delta, and $h(\beta_1, \ldots)$ is a
function of the indices that is 1 for a fake combination, and 0 for a real
combination.

In order to estimate the number of events contained within $\mathcal{\hat
T}^\eventfake$, from \eqnref{generalisedMatrixMethod}, that are tight, one sums
the appropriate component(s).  For example, a simple analysis selecting final
states with exactly two leptons might define
$\eventtight\equiv\objtight\objtight$ (\textit{i.e.} the number of tight
\emph{events} would now be denoted $n_T\equiv n_{tt}$), and all other
possibilities to be $\eventloose$.  In this case the $\mathcal{\hat
T}^\eventfake_{\objtight\objtight}$ component is the estimate of the number of
events that are both tight and fake.  For the completely general case with
events containing arbitrary numbers of leptons, additional terms and indices are
added as necessary to the equations in this section.

\subsubsection{Limit setting}
\label{subsubsec:matrixMethodLimits}
ATLAS and CMS analyses use the \cls method \cite{Mistlberger:2012rs} to place
an upper limit on the event rate (in the sense of the mean of a Poisson
distribution, which controls the appearance of events in a signal region) of new
physics processes. 

In the context of limit setting, the output from the matrix method is treated on
a par with those irreducible background components estimated from Monte Carlo
simulated (MC) samples.  Once the central value is estimated as described in
section \ref{subsec:matrixMethod}, uncertainties in the measured efficiencies,
as well as a statistical uncertainty, can be propagated in the usual way by
taking a derivative \cite{BIPM1995}.  The background mean $\bar{b}$ and
uncertainty $\sigma_b$ are fed into a joint likelihood for the signal and
background rates, $\mu$ and $b$, given the number of events observed in the
signal region $n_T$.  In the case with only one background source it takes
the form
\begin{align}
  \mathcal{L}(\mu, b | n_T) &= \text{Poiss}(n_T; \mu+b) \text{Gauss}(\bar{b}; b,
  \sigma_b).
  \label{eqn:likelihoodForSimpleBackground}
\end{align}

When setting the limit, the nuisance parameter $b$ is profiled away in the usual
way to form the test statistic $q_\mu$,
\begin{align}
  q_\mu &= \begin{cases}
  -2\ln\left( \dfrac
    {\mathcal{L}\left(\mu, \hat{\hat{b}} | n_T \right)}
    {\mathcal{L}\left(\hat\mu, \hat{b}  | n_T \right)}
  \right) & \mu > \hat\mu \\
  0 & \\
  \end{cases}.
\end{align}
Confidence intervals (\cls or \clsb) at some level can then be formed by
following the recipe outlined in \cite{Mistlberger:2012rs}.

\subsection{\methodB: An extended likelihood method}
\label{subsec:extendedLikelihoodMethod}
Whilst \methodA can suffer from under-coverage, as subsequently discussed in
\secref{frequentistLimitComparison}, this can largely be avoided for a purely data-driven
background if the \emph{full} likelihood, including all data used to make
the measurement, is used in the limit setting procedure. That is, one should use
\begin{align}
  \mathcal{L}(\mu, \bm{\theta} | n_t, n_l, n_{tt}, \ldots),
\end{align}
where $\bm\theta$ represents the set of nuisance parameters. If the leptons can
fall into one of several categories, quantities should be replaced with the
separate terms from \eqnref{nttExpansionWithCategories}. Each of these
quantities can be considered as an independent random variable with a Poisson
distribution. The means of these Poisson distributions will be denoted as
functions of the parameters e.g.
$\nu_{tt}^{\objcategory{1}\objcategory{1}}(\mu, \bm{\theta})$; the likelihood then
factorises and takes a form similar to \eqnref{likelihoodForSimpleBackground}
\begin{align}
  \mathcal{L}(\mu, \bm{\theta} | \ldots, n_{tt}^{\objcategory{1}\objcategory{1}},
    \ldots) = 
    \text{Poiss}(n_{tt}^{\objcategory{1}\objcategory{1}};
    \nu_{tt}^{\objcategory{1}\objcategory{1}}(\mu, \bm{\theta}))\cdots
    P(\bm{\bar\theta} | \bm{\theta}).
    \label{eqn:likelihoodPoissonProduct}
\end{align}
The final term represents constraints placed on the nuisance parameters by
external measurements.

\subsubsection{Choice of parameterisation}
The efficacy of any likelihood method depends on a sensible choice of parameterisation.
The parameterisation must completely describe how events from both
signal and background are expected to be divided between the different
event categories without over-parameterising. For example, one \emph{could} directly use
$\nu_{tt}^{\objcategory{1}\objcategory{1}}$ {\it etc.}~as the free parameters $\bm\theta$, but
this would remove all predictive power! 

Other researchers \cite{likelihoodMMCranmerComm} have investigated the
possibility of applying a method that uses a similar parameterisation to the
matrix method. This parameterisation uses the efficiencies described before, in addition to the rates
separated both by object category and real/fake-ness. Whilst this has an
advantage of making minimal assumptions about how a given background process
distributes itself between these categories, it does lead to a very large
parameter space. For example, even with three objects coming from only three possible
categories, there are already 80 such parameters (before considering
efficiencies). Since any form of prediction will require a maximisation of the
likelihood over this input parameter space, and since such global maximisations
become computationally more expensive as dimensionality increases, the authors
have chosen to use an alternative parameterisation.

\begin{figure*}
  \centering
  \includegraphics[width=\textwidth]{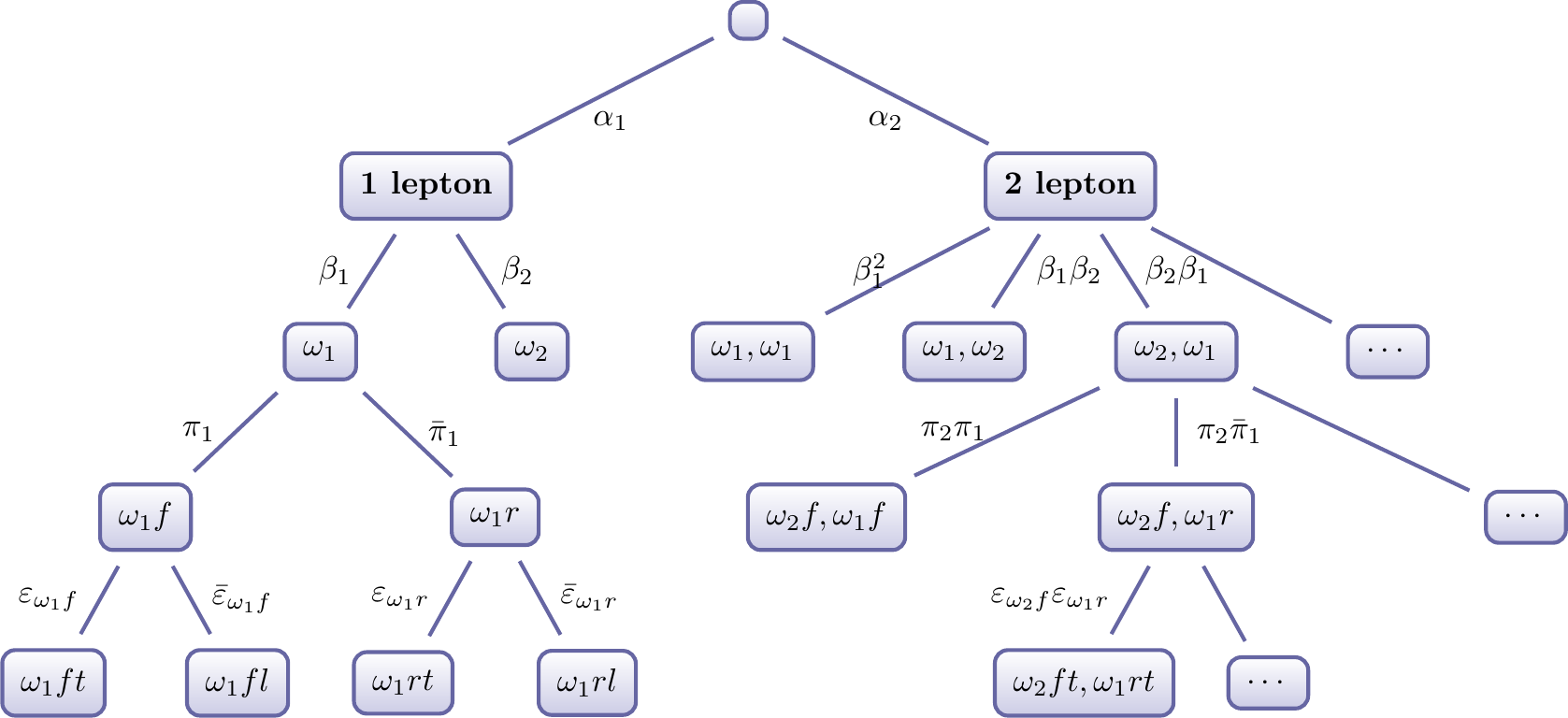}
  \caption{
    \label{fig:tomsTreeParams}
    The probability tree in this figure illustrates the model used to
    parameterise fake and real lepton production, as used in \methodsBandC.  The
    left-most branch is complete, the others are not as indicated by the
    presence of ``\ldots''. In general one could
    allow both for more lepton categories, as well as more leptons in the event.
    Note that $\sum_{\numlep=0}^{\numlep_\text{max}} \alpha_\numlep =
    \sum_{i=1}^\numobjcategories \beta_i = 1$, where $\numlep_\text{max}$ is the
    largest number of leptons that can be produced in a given event.
    Additionally, the abbreviation $\bar\pi_i = 1 - \pi_i$ is used.
}
\end{figure*}

Diagrammatically, the parameterisation used in this work is displayed in
\figref{tomsTreeParams}. For every event that is generated, it is first decided
how many leptons that event ought to contain. This is controlled by a set of
parameters $\{\alpha_\numlep\}$, each of which corresponds to the probability of
forming an event with \numlep leptons. As noted in the caption, these must sum to
1. For each lepton, a category \objcategory{i} is
assigned to it with probability $\beta_i$, and it is then further assigned to be
either \objreal with probability $\pi_i$ or \objfake with probability $1-\pi_i$.
Formally, $\beta_i \equiv P(\objcategory{i}|\objall)$ and $\pi_i \equiv
P(\objfake|\objcategory{i}\objall)$.
Efficiencies are then used in the usual way to select objects as being
\objtight or \objloose.

Using these terms, together with one extra non-negative parameter denoting the
mean of the Poisson distribution controlling the total production of tight
\emph{events},\footnote{One could alternatively use the overall production of
  \eventall events, however it is essential to have the rate of \eventtight
  events as a parameter for any signal component, since this is the quantity
upon which one wishes to place a limit.} one can compute the terms such as
$\nu_{tt}^{\objcategory{1}\objcategory{1}}$ in
\eqnref{likelihoodPoissonProduct}. It should be noted that one of these trees
must exist for every separate `component' that is being fitted -- that is, at
least one for the hypothesised signal process and one for the fake component of
the background, and then optionally one or more for other background components
that have been estimated using MC samples.

\subsection{\methodC: Maximum likelihood estimate}
\label{subsec:MLEMethod}
It is later found that \methodB\ is computationally intractable for more than
very simple systems. As such, we propose a third method that is found to keep
many of the desirable properties of \methodB, but also with a much reduced
computation time more similar to \methodA. This is achieved by using a simple
likelihood for limit setting, as in \methodA, but feeding it with the true \MLE\
fake rate.

To form an upper limit with \methodC, one does the following. Firstly, for the
observed data, maximise the likelihood expressed in
\eqnref{likelihoodPoissonProduct} for all nuisance parameters. As mentioned
in the discussion for \methodB, this likelihood should contain sufficient
parameters to describe the signal process, fake background, and any other
real backgrounds. The output from this that shall be used is the \MLE\ fake
rate with an estimated uncertainty. This uncertainty represents both an
uncertainty with which the efficiencies are known, as well as statistical
limitations of the observed data. It is estimated with the MINOS method
\cite{James:1975dr}, by taking the values of the fake rate where the minimum of
the negative log likelihood with respect to the \emph{remaining} parameters
increases by 0.5 from its minimum value. A limit is then placed using an
expression identical to that in \eqnref{likelihoodForSimpleBackground}, where
$\bar{b}$ and $\sigma_b$ take the aforementioned \MLE\ fake rate and
uncertainty.

\section{Comparisons using frequentist limits}
\label{sec:frequentistLimitComparison}
Using a toy event generator, written by the authors, datasets are produced using
the same method as that which is depicted in \figref{tomsTreeParams}, containing
a mixture of `fake' and `signal' events. For each of several configurations,
\numPseudoDatasets
independent datasets were formed using the generator. Each of these was
subsequently processed using \allmethods. In all cases the necessary
minimisation of a negative log likelihood was performed using the Minuit2 library
\cite{James:1975dr}. The result are 95\% \clsb and \cls upper
limits on the signal strength parameter.\footnote{The $p$-values used to compute
  \cls and \clsb are computed by performing pseudo-experiments, rather than using
  asymptotic methods \cite{2011EPJC...71.1554C}, since it is known that the
  latter are only a good approximation for scenarios with a large number of
events. In this work we focus on regions with low numbers of events.}

\blockoftext{}    

\subsection{Simple scenario -- two leptons, two categories}
Firstly, a configuration is used that produces events always with exactly two
leptons, each of which can be in one of two categories. There are separate
configurations for a signal process, which produces only real leptons
($\pi_1=\pi_2=0$), and a fake process which produces only fake leptons
($\pi_1=\pi_2=1$).  The full set of parameters can be found in
\appendixref{parameters} in \tableref{simpleExampleParameters}.
In each dataset, 100 events are produced using the tree in
\figref{tomsTreeParams}. As such the number of \eventtight events is
approximately the sum of two Poisson random variables; one representing the
signal component with mean 0.706, and another representing the fake background
with a mean of 1.94. 

\begin{figure}
  \centering
  \includegraphics[width=\columnwidth]{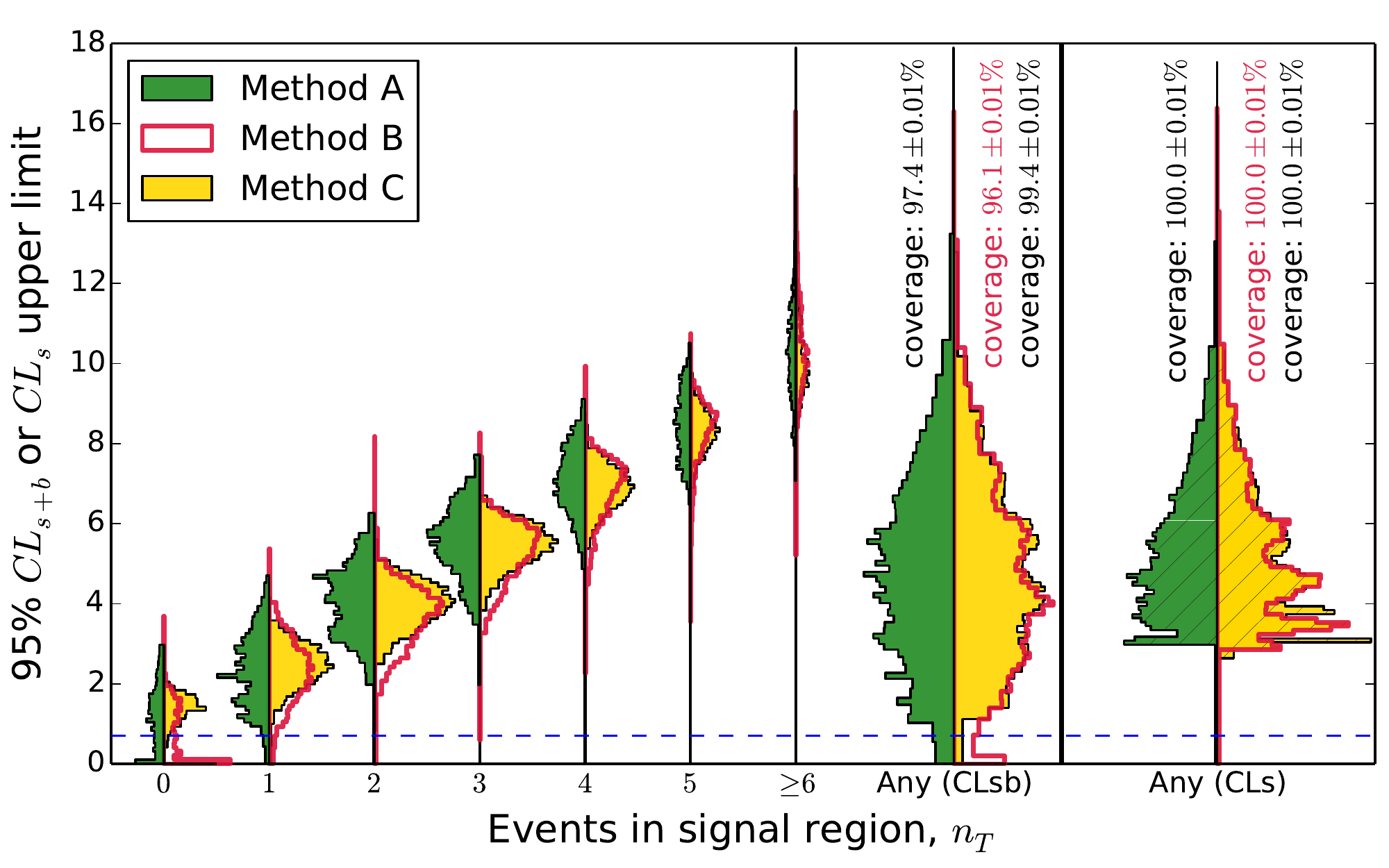}
  \caption{
    \label{fig:simpleExampleCLsLimits}
    In the {\it simple scenario} described in the text (two leptons in two categories) computations using \methodB\ are tractable.  
    This figure therefore compares \allmethods\ in that configuration.  The
    95\% \clsb and \cls upper limits on the rate of \eventtight signal events
    for each of \numPseudoDatasets independent toy datasets are histogrammed.
    For each `column', histograms are made for each of \allmethods\ and plotted
    back-to-back; that is A is plotted opposite to B and C, which overlap. The
    \clsb results are further divided into bins of observed \numTight; in all
    cases the area of each histogram is proportional to the number of toy
    datasets used to create it.  The dashed blue line indicates the true signal
    production rate, $\rateTightReal=0.706$. The coverage of the observed limits
    of this truth rate are noted for the overall \clsb and \cls results.
  }
\end{figure}

The \clsb and \cls limits from each of the \numPseudoDatasets generated datasets
is shown in \figref{simpleExampleCLsLimits}. The \clsb limit is shown to have
approximately correct coverage for \methodsAandB, but \methodC\ over-covers;
deviations from 95\% at this level can only be justified on the grounds of the
use of a profiled test statistic, rather than the full Neyman construction.
This is further considered for the next example.  There is also significant
over-coverage in the \cls limit, however this is expected due to the definition
$\cls = \frac{\clsb}{1-\clb}$. In low statistics regimes, often $(1-\clb)<1$,
meaning that $\cls>\clsb$ by a potentially significant margin.

Furthermore, a division of the \clsb limit according to the number of events
observed in the signal region, \numTight, is also shown in
\figref{simpleExampleCLsLimits}. From this figure, whilst it can be seen
that overall very similar limits are being placed by all three methods, in fact
\methodB tends to be most constraining due to its distributions showing
longer lower tails. \methodsBandC\ are together significantly more constraining than
\methodA\ (signified by shorter upper tails), and are quite similar to each other; this is
encouraging in justifying the use of \methodC\ as an approximation to \methodB.
\methodB is slightly more likely to place a tighter limit, as is to be
expected since it makes optimal use of all available information.

\begin{figure}
  \centering
  \includegraphics[width=\columnwidth]{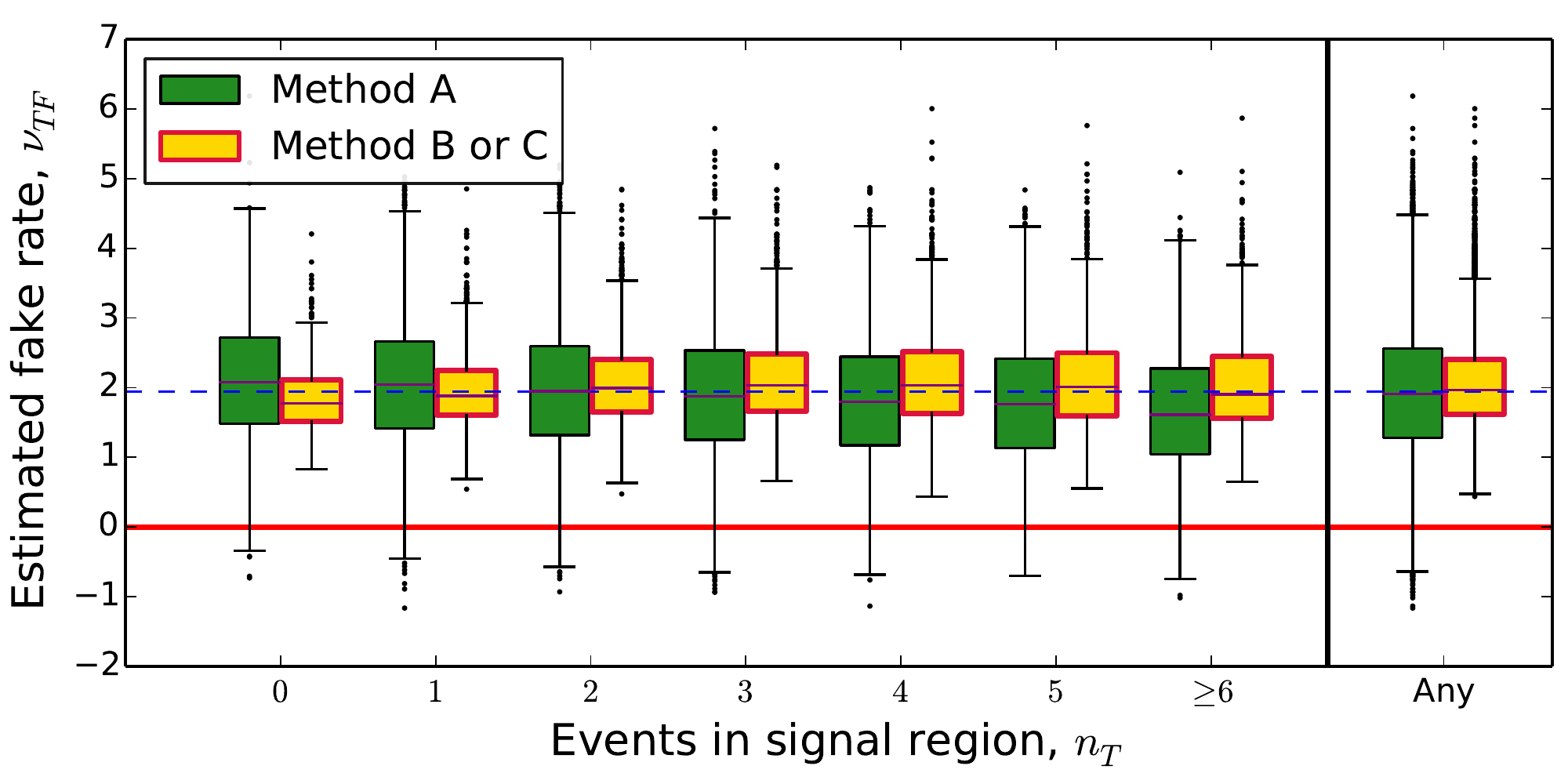}
  \caption{
    \label{fig:simpleExampleFakeRates}
    For the {\it simple scenario} described in the text (two leptons in two categories) in which 
    computations using \methodB\ remain tractable, the estimated fake rates for
    each of \numPseudoDatasets independent toy datasets are shown as a function
    of \numTight, comparing \methodsAandB\ with box plots. The fake rate from
    \methodC\ is by definition the same as that from \methodB.  The box plots indicate the
    median and lower \& upper quartiles with the box, while the whiskers extend
    to most extreme datum within $1.5\times$inter-quartile range of the nearest
    quartile; this corresponds to the $k=1.5$ case as described in
    \cite{boxplots}. Black dots are used to mark data points outside the range
    of the whiskers.  The dashed blue line marks the true value of
    $\rateTightFake=1.94$, and the red line delimits the unphysical
    $\rateTightFake<0$ region.
  }
\end{figure}

A further comparison that can be made is of the fake rate that is the output of
the matrix method in \methodA, against the \MLE\ of the fake rate obtained in
\methodsBandC ; this is shown in
\figref{simpleExampleFakeRates}. The spread in the plot demonstrates the
property that \methodA\ can predict a negative fake rate, as seen in a
significant portion of the generated datasets. It also shows that \methodB\
produces fake rates that cluster more closely around the true value, even at low
\numTight.

\subsection{Harder scenario -- two leptons, eight categories}
The simple scenario above has been extended to use eight categories instead of
two. As per the parameterisation in \figref{tomsTreeParams}, this involves the addition of 24
extra parameters -- twelve each for the signal and fake background from the
addition of six $\beta$ and six $\pi$ terms. The full set of parameters can be
found in \appendixref{parameters} in \tableref{harderExampleParameters}.
As before, 100 events were generated in each dataset, corresponding to a signal
rate of 0.748 and a fake background rate of 2.77.

It was found that the increase in parameter space dimensionality was sufficient
to increase the computation time for the likelihood maximisation to such an extent that
producing limits with \methodB became infeasible using the resources at the
authors' disposal; as such only results from \methodA\ and \methodC\ could be
computed.

\begin{figure}
  \centering
  \includegraphics[width=\columnwidth]{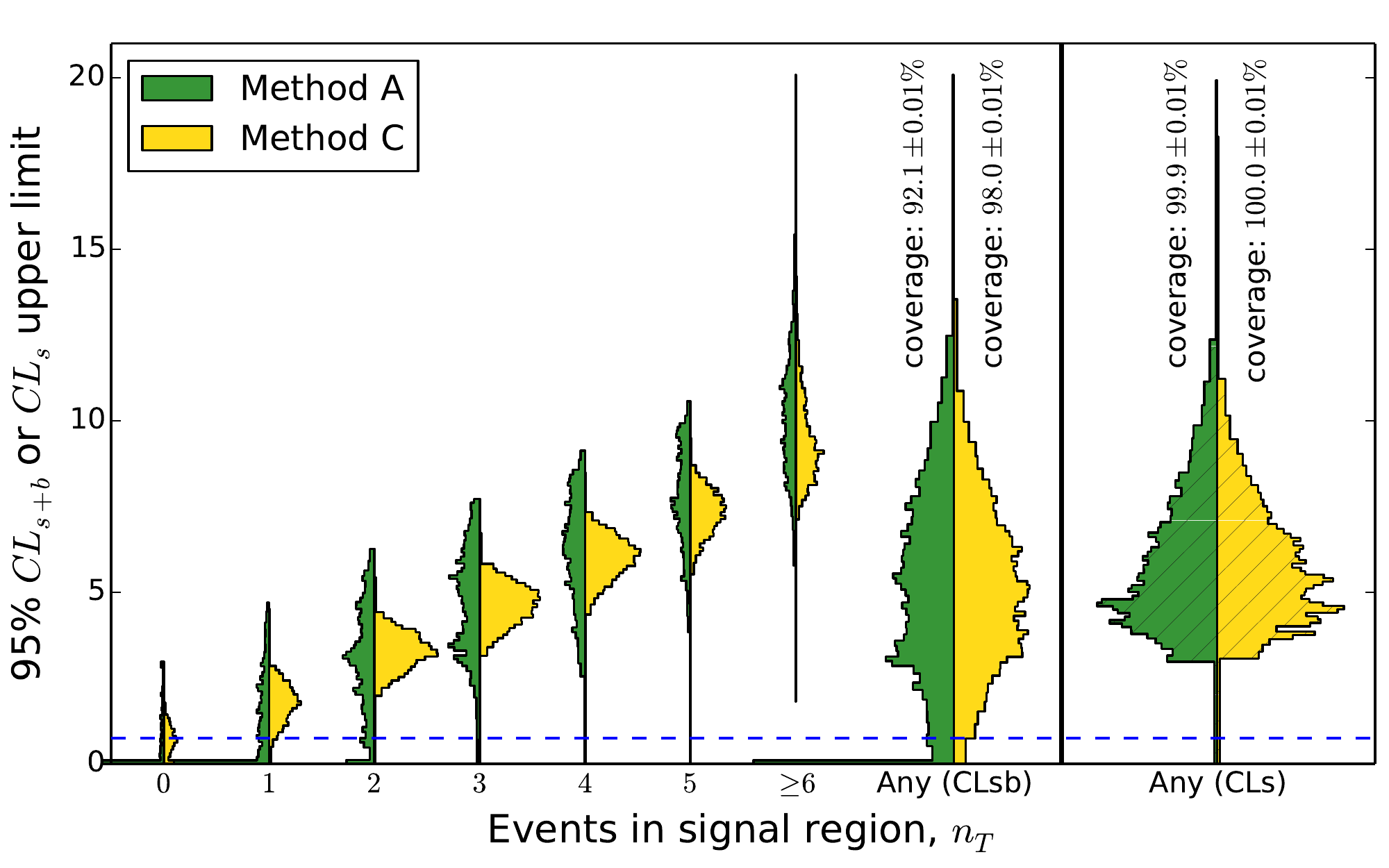}
  \caption{
    \label{fig:harderExampleCLsLimits}
    In the {\it harder scenario} described in the text (two leptons, eight categories) 
 computations using \methodB\ are no longer tractable.  This figure therefore compares the only two usable Methods (A and C) in this more challenging case.  The 95\% \clsb and \cls
    upper limits on the rate of \eventtight signal events for each of
    \numPseudoDatasets independent toy datasets are histogrammed. For each
    `column', histograms are made for each of Methods A and C and plotted
    back-to-back.  The \clsb results are further divided into bins of observed
    \numTight; in all cases the area of each histogram is proportional to the
    number of toy datasets used to create it. The dashed blue line indicates the
    true signal production rate, $\rateTightReal=0.748$. The coverage of the
    observed limits of this truth rate are noted for the overall \clsb and \cls
    results.
  }
\end{figure}
\begin{figure}
  \centering
  \includegraphics[width=\columnwidth]{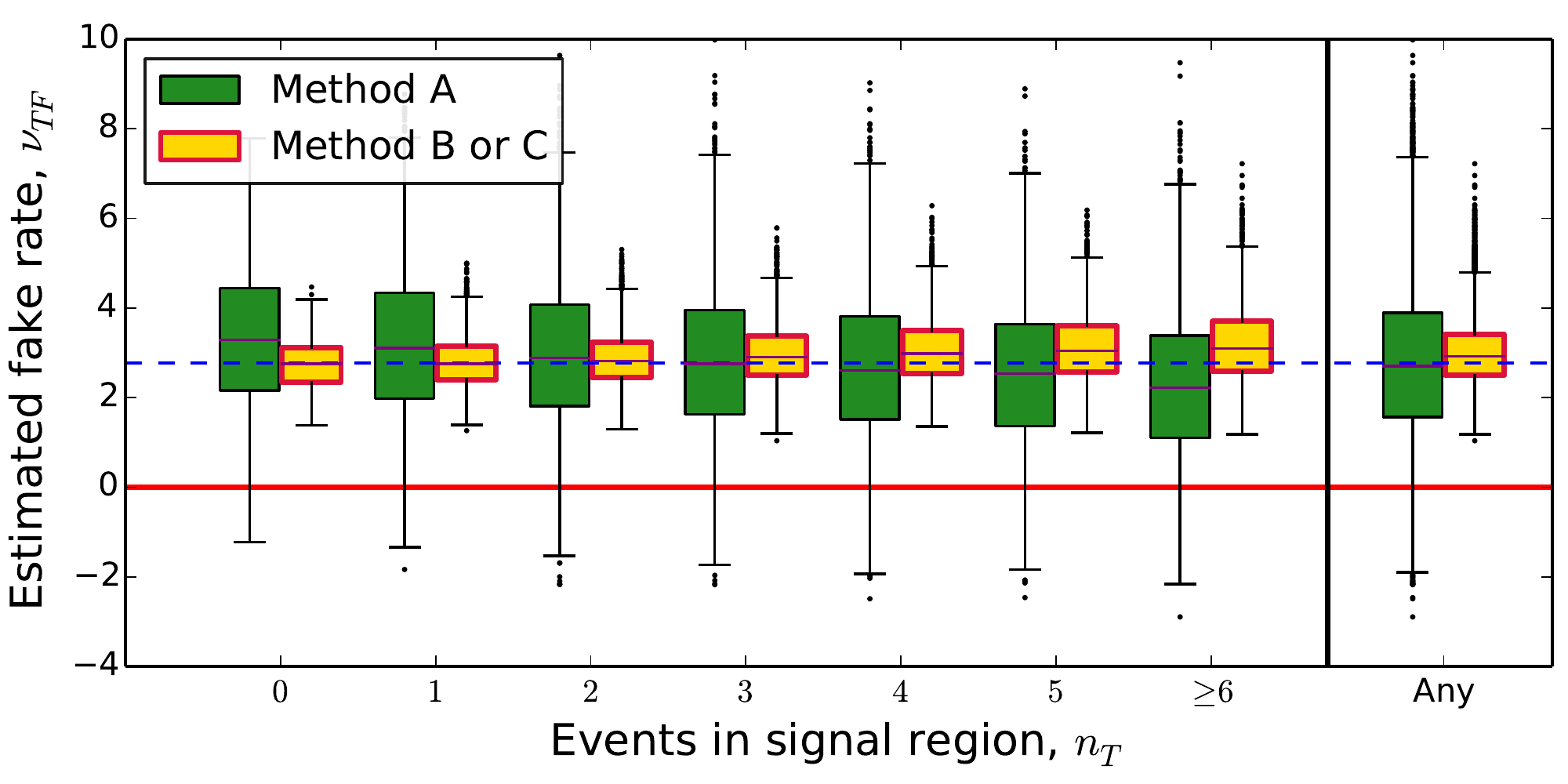}
  \caption{
    \label{fig:harderExampleFakeRates}
    For the {\it harder scenario} described in the text (two leptons, eight categories)
     the fake rates estimated by Methods A and C 
    for each of \numPseudoDatasets independent toy datasets are shown as a function of
    \numTight in the form of box plots, similarly to
    \figref{simpleExampleFakeRates}. The dashed blue line marks the true
    value of $\rateTightFake=2.77$, and the red line delimits the unphysical
    $\rateTightFake<0$ region.
  }
\end{figure}

\Figref{harderExampleFakeRates} shows that the \MLE\ fake rate for \methodC is
much more tightly constrained around the true value than the \methodA estimate;
moreover \methodA gives even more significant deviations into negative values
than with the simple scenario. Furthermore, as \numTight increases, the median fake
rate from \methodA decreases slightly, whereas that from \methodC is stable for
low event counts, only increasing slightly for larger \numTight; the \methodC
behaviour seems more desirable here. Secondly, \figref{harderExampleCLsLimits}
shows that the \clsb limits derived in \methodA suffer from under-coverage;\footnote{At least within particle physics, under-coverage is typically considered a worse crime than over-coverage, as the related limits are then not conservative.} the
upper limit only bounds the true rate 92\% of the time rather than the expected
95\%.  Finally the upper tails of the \clsb limit are significantly more
pronounced in \methodA than in \methodC, as can be seen when the limits are
separated by \numTight, as also included in \figref{harderExampleCLsLimits}.

\section{Conclusions}
We have described the matrix method, used
in many ATLAS and CMS analyses to estimate fake leptonic backgrounds, more completely than we have seen elsewhere. We have shown 
that it (\methodA) produces a \MLE\ fake rate under
a restrictive set of conditions, but that these are rarely met in practice.  We have shown that it has a number of undesired properties which result from its heuristic definition: (i) it can give
physically meaningless results (predict negative fake rates), (ii) its fake rate estimates show an undesired bias as a function of the number of tight events
 in the signal region (as seen by the slope observed in the run of green boxes of Figures~\ref{fig:simpleExampleFakeRates} and \ref{fig:harderExampleFakeRates}), and (iii) the limits it sets on fake rates are significantly more variable than those from better methods (as seen by the increased vertical extent of the
 \methodA\ histograms in Figures~\ref{fig:simpleExampleCLsLimits} and
 \ref{fig:harderExampleCLsLimits} compared to those of \methodsBandC).   We noted that, within the constraints of the frequentist profile-likelihood based framework considered, one cannot hope to constrain fake rates much better than \methodB.
  However, we saw that the computational overheads of \methodB\ precluded its use in all but the simplest of cases.  Finally, we showed that it was possible to find a third approach, \methodC, which is computationally of similar complexity to \methodA and, though to some extent also heuristic in its definition, nonetheless reproduces much more closely the fake rate estimates of \methodB.  \methodC, in contrast to the matrix method: (i) gives only physically meaningful results (predicts positive fake rates), (ii) its fake rate estimates are unbiased as a function of the number of tight events in the signal region (as seen by the lack of slope across the 
yellow boxes of  Figures~\ref{fig:simpleExampleFakeRates} and \ref{fig:harderExampleFakeRates}), and (iii) the limits it sets on fake rates are significantly less variable than those of \methodA while being very close to those seen in the optimal Method B (again see Figures~\ref{fig:simpleExampleCLsLimits} and \ref{fig:harderExampleCLsLimits}).
The improvements seen in \methodC over \methodA are particularly notable in
signal regions having few events.

A possible advantage of \methodsBandC\ not explored in this paper is afforded by
their ability, if desired, to encapsulate background processes which can
contribute by both real and fake events \textit{e.g.} in different decay modes
by use of different parameter trees. Further to this, it may be found that
measurements in additional regions could constrain some of these parameters,
much like the efficiencies are already constrained. Whilst speculative, further
research in this area has the potential to provide even greater benefits over
the original matrix method.

\acknowledgments
The authors acknowledge funding and support from the Science and Technology
Facilities Research Council (STFC) and Peterhouse, and wish to thank Kyle
Cranmer for useful discussions relating to the possibility of use of methods
similar to \methodB.

\appendix

\section{Origin of matrix method approximation}
\label{appendix:origin} 
It was stated earlier that, under appropriate conditions, $\hat n_R$ and $\hat
n_F$ are maximum likelihood estimators for $\expn{R}\equiv\expectation{\numReal
| \numTight,\numLoose}$ and $\expn{F}\equiv\expectation{\numFake |
\numTight,\numLoose}$.  This result, together with its limitations, is now
presented.

\subsection{Single lepton and single category}
We shall firstly demonstrate this approximation in the simplified case with just
a single lepton and category. The corresponding fully general derivation follows
in the next section, but the logic it follows is largely the same.

When considering a likelihood as a product of Poisson terms as in
\eqnref{likelihoodForSimpleBackground}, and neglecting the Gaussian terms
involving the efficiencies, the negative log likelihood for the term arising
from the background component will be
\begin{align}
  \nll &= \sum_{\mathcal{T}\in\{T,L\}}
  \left(
      \tensRateTL
      -
      \tensTL \ln
        \tensRateTL
    \right).
  \label{eqn:nllForMMParamSimple}
\end{align}
Here we sum over two constraints, one from tight events and the other from
loose. The means of Poisson distributions are denoted by \tensRateTL. 
From \eqnsref{basicMatrixMethodForRates}{twoLeptonMM}, one finds
\begin{align}
  \tensRateTL &=
    \sum_{\mathcal{R}\in\{R,F\}}
      \phi_{\mathcal{T}\mathcal{R}}
      \tensRateRF,
  \label{eqn:generalisedExpRFtoExpTLSimple}
\end{align}
where \tensRateRF\ denotes the means of the Poisson distributions from which the
numbers of real and fake events are drawn, and $\phi$ is the matrix of
efficiencies with indices $\mathcal{T}$ and $\mathcal{R}$ referring to
tight/loose and real/fake properties respectively.

One can now differentiate \eqnref{nllForMMParamSimple} with respect to
$\tensRateRF,\forall\mathcal{R}\in\{R,F\}$, using the identity in
\eqnref{generalisedExpRFtoExpTLSimple}, and find the \MLE\ values for the rates,
denoted \tensRateMLETL.  In order to locate the minimum of the
negative log likelihood, one sets these derivatives to 0, yielding
\begin{align}
  \sum_{\mathcal{T}\in\{T,L\}}
  \left(
    1 - \frac{\tensTL}{\tensRateMLETL}
  \right)
  \phi_{\mathcal{T}\mathcal{R}} &= 0,
  \quad
  \forall\mathcal{R}\in\{R,F\}.
  \label{eqn:nllMinimumForMMParamSimple}
\end{align}
These are satisfied if $\tensRateMLETL = \tensTL\ \forall\mathcal{T}\in\{T,L\}$, the result of
which being that, upon inversion, \eqnref{generalisedExpRFtoExpTLSimple} will look like
\begin{align}
  \tensRateMLERF =
  \sum_{\mathcal{T}\in\{T,L\}}
  \phi^{-1}_{\mathcal{R}\mathcal{T}}
     \tensTL =
     \tensRFHat,
   \label{eqn:mmIsSometimesMLESimple}
\end{align}
analogously to \eqnref{basicMMInversionExact}.

Whilst this is a valid operation for the problem as stated above, it should be
noted that the minimum of \nll is represented by
\eqnref{nllMinimumForMMParamSimple} \emph{only} when the components of
$\tensRateMLERF$ are $>0$.  

To conclude, \eqnref{mmIsSometimesMLESimple} shows how the matrix method estimator
\tensRFHat\ is identical to the \MLE\ values \tensRateMLERF\ in the simplified
single-lepton matrix method, \emph{if} the condition above is met.

\subsection{Multiple leptons and multiple categories}
When considering a likelihood as a product of Poisson terms as in
\eqnref{likelihoodForSimpleBackground}, and neglecting the Gaussian terms
involving the efficiencies, the negative log likelihood for the term arising
from the background component will be
\begin{align}
  \nll &= \sum_{\bm{\omega},\bm{\beta}} 
    \left(
      \tensRateTL^{\bm{\omega}}_{\bm{\beta}}
      -
      \tensTL^{\bm{\omega}}_{\bm{\beta}} \ln
        \tensRateTL^{\bm{\omega}}_{\bm{\beta}}
    \right),
  \label{eqn:nllForMMParam}
\end{align}
where for a set of \numlep leptons the categories and tight/looseness
information are compacted into vectors $\bm{\omega}$ and $\bm{\beta}$ of length
\numlep respectively. Note also that the means of Poisson distributions are
denoted in the general notation by \textit{e.g.} \tensRateTL. 
From \eqnsref{basicMatrixMethodForRates}{twoLeptonMM}, one finds
\begin{align}
   \tensRateTL^{\bm{\omega}}_{\bm{\beta}} =
   \sum_{\bm{\alpha'}} 
       \fakePhi{\omega}{\beta}{\alpha'}
       \tensRateRF^{\bm{\omega}}_{\bm{\alpha'}},
   \fakePhi{\omega}{\beta}{\alpha'} = 
      \fakephi{i_1}{\beta_1}{\alpha'_1}
      \fakephi{i_2}{\beta_2}{\alpha'_2}
      \ \cdots,
  \label{eqn:generalisedExpRFtoExpTL}
\end{align}
where $\bm{\alpha'}$ is a vector representing whether each lepton is real or
fake.

One can now differentiate \eqnref{nllForMMParam} with respect to
$\tensRateRF^{\bm{\omega}}_{\bm{\alpha}},\ \forall
\bm{\omega},\bm{\alpha}$, using the identity in
\eqnref{generalisedExpRFtoExpTL}, and find the \MLE\ values for the rates,
denoted \textit{e.g.}~\tensRateMLETL.  In order to locate the minimum of the
negative log likelihood, one sets all these derivatives to 0, yielding
\begin{align}
  \sum_{\bm{\beta'}} 
  \left(
    1 - \frac{\tensTL^{\bm{\omega}}_{\bm{\beta'}}}
      {\tensRateMLETL^{\bm{\omega}}_{\bm{\beta'}}}
  \right)
  \fakePhi{\omega}{\alpha}{\beta'} &= 0,
  \quad
  \forall
  \bm{\omega},\bm{\alpha}.
  \label{eqn:nllMinimumForMMParam}
\end{align}
These are satisfied if $\tensRateMLETL^{\bm{\omega}}_{\bm{\beta'}} =
\tensTL^{\bm{\omega}}_{\bm{\beta'}}\ \forall \bm{\beta'}$, the result of
which being that, upon inversion, \eqnref{generalisedExpRFtoExpTL} will look like
\begin{align}
  \tensRateMLERF^{\bm{\omega}}_{\bm{\alpha}} =
  \sum_{\bm{\beta'}} 
     \fakeinvPhi{\omega}{\beta'}{\alpha}
     \tensTL^{\bm{\omega}}_{\bm{\beta'}} =
     \tensRFHat^{\bm{\omega}}_{\bm{\alpha}},
   \label{eqn:mmIsSometimesMLE}
\end{align}
analogously to \eqnref{basicMMInversionExact}.

Whilst this is a valid operation for the problem as stated above, it should be
noted that the minimum of \nll is represented by \eqnref{nllMinimumForMMParam}
\emph{only} when the components of
$\tensRateMLERF^{\bm{\omega}}_{\bm{\alpha}}$ are $>0$. Incidentally, is also only
useful in the case where the components of
$\tensRateMLERF^{\bm{\omega}}_{\bm{\alpha}}$ are readily assigned to
either signal plus other `real' backgrounds (those typically estimated from MC
samples) and the fake background. 

To conclude, \eqnref{mmIsSometimesMLE} shows how the matrix method estimator
\tensRFHat\ is identical to the \MLE\ values \tensRateMLERF\ in the fully
generalised matrix method, \emph{if} the conditions above are met.

\section{Tables of parameters}
\label{appendix:parameters} 
Parameters used to configure the toy generator may be found in the following tables: 

\begin{table}[h]
  \centering
  \renewcommand{\arraystretch}{1.2}
  \begin{tabular}{c@{\hskip 8\tabcolsep}ccc@{\hskip 8\tabcolsep}ccc@{\hskip 8\tabcolsep}cc}
    \toprule
    \textbf{Object}& \multicolumn{3}{@{}c@{\hskip 8\tabcolsep}}{\textbf{Signal}} & \multicolumn{3}{@{}c@{\hskip 8\tabcolsep}}{\textbf{Background}} && \\
    \textbf{category} & 
      $\nu_\eventall$ & $\beta$ & $\pi$ & 
      $\nu_\eventall$ & $\beta$ & $\pi$ & 
      $\effReal$ & $\effFake$ \\
    \midrule
    \objcategory{1} & 0.01 & 0.6 & 0 & 0.99 & 0.6 & 1 & 0.8 & 0.1 \\
    \objcategory{2} & --   & 0.4 & 0 & --   & 0.4 & 1 & 0.9 & 0.2 \\
    \bottomrule
  \end{tabular}
  \caption{
    \label{table:simpleExampleParameters}
    Parameters controlling the simple scenario with exactly two leptons, and two
    categories for each lepton. The parameters are as described in
    \figref{tomsTreeParams}, however $\alpha_2=1$ and $\alpha_i=0\ \forall
    i\neq2$.  The overall production rate of events is $\nu_\eventall$, each one
    of which is filtered through the decision tree.  Components marked with a
    `--' are not applicable in the context.
  }
\end{table}

\begin{table}[h]
  \centering
  \renewcommand{\arraystretch}{1.2}
  \begin{tabular}{c@{\hskip 8\tabcolsep}ccc@{\hskip 8\tabcolsep}ccc@{\hskip 8\tabcolsep}cc}
    \toprule
    \textbf{Object}& \multicolumn{3}{@{}c@{\hskip 8\tabcolsep}}{\textbf{Signal}} & \multicolumn{3}{@{}c@{\hskip 8\tabcolsep}}{\textbf{Background}} && \\
    \textbf{category} & 
      $\nu_\eventall$ & $\beta$ & $\pi$ & 
      $\nu_\eventall$ & $\beta$ & $\pi$ & 
      $\effReal$ & $\effFake$ \\
    \midrule
    \objcategory{1} & 0.01 & 0.086 & 0 & 0.99 & 0.184 & 1 & 0.8 & 0.1 \\
    \objcategory{2} & --   & 0.143 & 0 & --   & 0.008 & 1 & 0.8 & 0.2 \\
    \objcategory{3} & --   & 0.110 & 0 & --   & 0.182 & 1 & 0.8 & 0.1 \\
    \objcategory{4} & --   & 0.010 & 0 & --   & 0.123 & 1 & 0.8 & 0.3 \\
    \objcategory{5} & --   & 0.092 & 0 & --   & 0.102 & 1 & 0.9 & 0.2 \\
    \objcategory{6} & --   & 0.284 & 0 & --   & 0.081 & 1 & 0.9 & 0.1 \\
    \objcategory{7} & --   & 0.245 & 0 & --   & 0.106 & 1 & 0.9 & 0.4 \\
    \objcategory{8} & --   & 0.030 & 0 & --   & 0.214 & 1 & 0.9 & 0.1 \\
    \bottomrule
  \end{tabular}
  \caption{
    \label{table:harderExampleParameters}
    Parameters controlling the simple scenario with exactly two leptons, and
    eight categories for each lepton. The parameters are as described in
    \figref{tomsTreeParams}, however $\alpha_2=1$ and $\alpha_i=0\ \forall
    i\neq2$.  The overall production rate of events is $\nu_\eventall$, each one
    of which is filtered through the decision tree.  Components marked with a
    `--' are not applicable in the context.
  }
\end{table}

\clearpage

\bibliographystyle{JHEP}
\bibliography{paper,bibtexentriesATLAS,bibtexentriesCMS}

\providecommand{\href}[2]{#2}\begingroup\raggedright\begin{thebibliography}{10}

\bibitem{ATLAS-CONF-2013-061}
{\bf ATLAS} Collaboration, {\it {Search for strong production of supersymmetric
  particles in final states with missing transverse momentum and at least three
  b-jets using 20.1 fb−1 of pp collisions at sqrt(s) = 8 TeV with the ATLAS
  Detector.}},  Tech. Rep. ATLAS-CONF-2013-061, CERN, Geneva, Jun, 2013.

\bibitem{ATLAS-CONF-2010-087}
{\bf ATLAS} Collaboration, {\it {Background studies for top-pair production in
  lepton plus jets final states in $\sqrt{s}=7 TeV$ ATLAS data}},  Tech. Rep.
  ATLAS-CONF-2010-087, CERN, Geneva, Oct, 2010.

\bibitem{atlasMMCitation0}
{\bf ATLAS} Collaboration, {\it {Search for supersymmetry using final states
  with one lepton, jets, and missing transverse momentum with the ATLAS
  detector in $\sqrt{s}=7$ TeV $pp$}},  {\em Phys.Rev.Lett.} {\bf 106} (2011)
  131802, [\href{http://arxiv.org/abs/1102.2357}{{\tt arXiv:1102.2357}}].

\bibitem{atlasMMCitation1}
{\bf ATLAS} Collaboration, {\it {Search for an excess of events with an
  identical flavour lepton pair and significant missing transverse momentum in
  $\sqrt{s}=7$ TeV proton-proton collisions with the ATLAS detector}},  {\em
  Eur.Phys.J.} {\bf C71} (2011) 1647,
  [\href{http://arxiv.org/abs/1103.6208}{{\tt arXiv:1103.6208}}].

\bibitem{atlasMMCitation2}
{\bf ATLAS} Collaboration, {\it {Search for supersymmetric particles in events
  with lepton pairs and large missing transverse momentum in $\sqrt{s}=7$ TeV
  proton-proton collisions with the ATLAS experiment}},  {\em Eur.Phys.J.} {\bf
  C71} (2011) 1682, [\href{http://arxiv.org/abs/1103.6214}{{\tt
  arXiv:1103.6214}}].

\bibitem{atlasMMCitation3}
{\bf ATLAS} Collaboration, {\it {Search for pair production of first or second
  generation leptoquarks in proton-proton collisions at $\sqrt{s}=7$ TeV using
  the ATLAS detector at the LHC}},  {\em Phys.Rev.} {\bf D83} (2011) 112006,
  [\href{http://arxiv.org/abs/1104.4481}{{\tt arXiv:1104.4481}}].

\bibitem{atlasMMCitation4}
{\bf ATLAS} Collaboration, {\it {Measurement of the cross section for the
  production of a $W$ boson in association with $b^-$ jets in $pp$ collisions
  at $\sqrt{s}=7$ TeV with the ATLAS detector}},  {\em Phys.Lett.} {\bf B707}
  (2012) 418--437, [\href{http://arxiv.org/abs/1109.1470}{{\tt
  arXiv:1109.1470}}].

\bibitem{atlasMMCitation5}
{\bf ATLAS} Collaboration, {\it {Search for anomalous production of prompt
  like-sign muon pairs and constraints on physics beyond the Standard Model
  with the ATLAS detector}},  {\em Phys.Rev.} {\bf D85} (2012) 032004,
  [\href{http://arxiv.org/abs/1201.1091}{{\tt arXiv:1201.1091}}].

\bibitem{atlasMMCitation6}
{\bf ATLAS} Collaboration, {\it {Search for pair-produced heavy quarks decaying
  to Wq in the two-lepton channel at $\sqrt{s}=7$ TeV with the ATLAS
  detector}},  {\em Phys.Rev.} {\bf D86} (2012) 012007,
  [\href{http://arxiv.org/abs/1202.3389}{{\tt arXiv:1202.3389}}].

\bibitem{atlasMMCitation7}
{\bf ATLAS} Collaboration, {\it {Observation of spin correlation in $t \bar{t}$
  events from pp collisions at sqrt(s) = 7 TeV using the ATLAS detector}},
  {\em Phys.Rev.Lett.} {\bf 108} (2012) 212001,
  [\href{http://arxiv.org/abs/1203.4081}{{\tt arXiv:1203.4081}}].

\bibitem{atlasMMCitation8}
{\bf ATLAS} Collaboration, {\it {A search for $t\bar{t}$ resonances with the
  ATLAS detector in 2.05 fb$^{-1}$ of proton-proton collisions at $\sqrt{s}=7$
  TeV}},  {\em Eur.Phys.J.} {\bf C72} (2012) 2083,
  [\href{http://arxiv.org/abs/1205.5371}{{\tt arXiv:1205.5371}}].

\bibitem{atlasMMCitation9}
{\bf ATLAS} Collaboration, {\it {A search for $t\bar{t}$ resonances in
  lepton+jets events with highly boosted top quarks collected in $pp$
  collisions at $\sqrt{s} = 7$ TeV with the ATLAS detector}},  {\em JHEP} {\bf
  1209} (2012) 041, [\href{http://arxiv.org/abs/1207.2409}{{\tt
  arXiv:1207.2409}}].

\bibitem{atlasMMCitation10}
{\bf ATLAS} Collaboration, {\it {ATLAS search for a heavy gauge boson decaying
  to a charged lepton and a neutrino in $pp$ collisions at $\sqrt{s}=7$ TeV}},
  {\em Eur.Phys.J.} {\bf C72} (2012) 2241,
  [\href{http://arxiv.org/abs/1209.4446}{{\tt arXiv:1209.4446}}].

\bibitem{atlasMMCitation11}
{\bf ATLAS} Collaboration, {\it {Search for resonant top plus jet production in
  $t\bar{t}$ + jets events with the ATLAS detector in $pp$ collisions at
  $\sqrt{s}=7$ TeV}},  {\em Phys.Rev.} {\bf D86} (2012) 091103,
  [\href{http://arxiv.org/abs/1209.6593}{{\tt arXiv:1209.6593}}].

\bibitem{atlasMMCitation12}
{\bf ATLAS} Collaboration, {\it {Search for Supersymmetry in Events with Large
  Missing Transverse Momentum, Jets, and at Least One Tau Lepton in 7 TeV
  Proton-Proton Collision Data with the ATLAS Detector}},  {\em Eur.Phys.J.}
  {\bf C72} (2012) 2215, [\href{http://arxiv.org/abs/1210.1314}{{\tt
  arXiv:1210.1314}}].

\bibitem{atlasMMCitation13}
{\bf ATLAS} Collaboration, {\it {Measurement of $ZZ$ production in $pp$
  collisions at $\sqrt{s}=7$ TeV and limits on anomalous $ZZZ$ and $ZZ\gamma$
  couplings with the ATLAS detector}},  {\em JHEP} {\bf 1303} (2013) 128,
  [\href{http://arxiv.org/abs/1211.6096}{{\tt arXiv:1211.6096}}].

\bibitem{atlasMMCitation14}
{\bf ATLAS} Collaboration, {\it {Multi-channel search for squarks and gluinos
  in $\sqrt{s}=7$ TeV $pp$ collisions with the ATLAS detector}},  {\em
  Eur.Phys.J.} {\bf C73} (2013) 2362,
  [\href{http://arxiv.org/abs/1212.6149}{{\tt arXiv:1212.6149}}].

\bibitem{atlasMMCitation15}
{\bf ATLAS} Collaboration, {\it {Search for $t\bar t$ resonances in the lepton
  plus jets final state with ATLAS using 4.7 fb$^{-1}$ of $pp$ collisions at
  $\sqrt{s} = 7$ TeV}},  {\em Phys.Rev.} {\bf D88} (2013), no.~1 012004,
  [\href{http://arxiv.org/abs/1305.2756}{{\tt arXiv:1305.2756}}].

\bibitem{atlasMMCitation16}
{\bf ATLAS} Collaboration, {\it {Measurement of Top Quark Polarization in
  Top-Antitop Events from Proton-Proton Collisions at $\sqrt{s}$ = 7  TeV
  Using the ATLAS Detector}},  {\em Phys.Rev.Lett.} {\bf 111} (2013), no.~23
  232002, [\href{http://arxiv.org/abs/1307.6511}{{\tt arXiv:1307.6511}}].

\bibitem{atlasMMCitation17}
{\bf ATLAS} Collaboration, {\it {Measurement of the top quark pair production
  charge asymmetry in proton-proton collisions at $\sqrt{s}$ = 7 TeV using the
  ATLAS detector}},  {\em JHEP} {\bf 1402} (2014) 107,
  [\href{http://arxiv.org/abs/1311.6724}{{\tt arXiv:1311.6724}}].

\bibitem{atlasMMCitation18}
{\bf ATLAS} Collaboration, {\it {Search for a Multi-Higgs Boson Cascade in
  $W^+W^− b\bar{b}$ events with the ATLAS detector in pp collisions at √s =
  8 TeV}},  {\em Phys.Rev.} {\bf D89} (2014) 032002,
  [\href{http://arxiv.org/abs/1312.1956}{{\tt arXiv:1312.1956}}].

\bibitem{atlasMMCitation19}
{\bf ATLAS} Collaboration, {\it {Measurement of the production of a $W$ boson
  in association with a charm quark in $pp$ collisions at $\sqrt{s} =$ 7 TeV
  with the ATLAS detector}},  {\em JHEP} {\bf 1405} (2014) 068,
  [\href{http://arxiv.org/abs/1402.6263}{{\tt arXiv:1402.6263}}].

\bibitem{atlasMMCitation20}
{\bf ATLAS} Collaboration, {\it {Search for direct top-squark pair production
  in final states with two leptons in pp collisions at $\sqrt{s} =$ 8TeV with
  the ATLAS detector}},  {\em JHEP} {\bf 1406} (2014) 124,
  [\href{http://arxiv.org/abs/1403.4853}{{\tt arXiv:1403.4853}}].

\bibitem{atlasMMCitation21}
{\bf ATLAS} Collaboration, {\it {Search for direct top squark pair production
  in events with a Z boson, b-jets and missing transverse momentum in
  $\sqrt{s}$=8 TeV pp collisions with the ATLAS detector}},  {\em Eur.Phys.J.}
  {\bf C74} (2014) 2883, [\href{http://arxiv.org/abs/1403.5222}{{\tt
  arXiv:1403.5222}}].

\bibitem{atlasMMCitation22}
{\bf ATLAS} Collaboration, {\it {Search for direct production of charginos,
  neutralinos and sleptons in final states with two leptons and missing
  transverse momentum in $pp$ collisions at $\sqrt{s} =$ 8 TeV with the ATLAS
  detector}},  {\em JHEP} {\bf 1405} (2014) 071,
  [\href{http://arxiv.org/abs/1403.5294}{{\tt arXiv:1403.5294}}].

\bibitem{atlasMMCitation23}
{\bf ATLAS} Collaboration, {\it {Search for supersymmetry at $\sqrt{s}$=8 TeV
  in final states with jets and two same-sign leptons or three leptons with the
  ATLAS detector}},  {\em JHEP} {\bf 1406} (2014) 035,
  [\href{http://arxiv.org/abs/1404.2500}{{\tt arXiv:1404.2500}}].

\bibitem{atlasMMCitation24}
{\bf ATLAS} Collaboration, {\it {Search for microscopic black holes and string
  balls in final states with leptons and jets with the ATLAS detector at
  $\sqrt{s}$ = 8 TeV}},  \href{http://arxiv.org/abs/1405.4254}{{\tt
  arXiv:1405.4254}}.

\bibitem{atlasMMCitation25}
{\bf ATLAS} Collaboration, {\it {Search for top squark pair production in final
  states with one isolated lepton, jets, and missing transverse momentum in
  $\sqrt{s}=$ 8 TeV pp collisions with the ATLAS detector}},
  \href{http://arxiv.org/abs/1407.0583}{{\tt arXiv:1407.0583}}.

\bibitem{atlasMMCitation26}
{\bf ATLAS} Collaboration, {\it {Search for strong production of supersymmetric
  particles in final states with missing transverse momentum and at least three
  b-jets at $\sqrt{s} =$ 8 TeV proton-proton collisions with the ATLAS
  detector}},  \href{http://arxiv.org/abs/1407.0600}{{\tt arXiv:1407.0600}}.

\bibitem{atlasMMCitation27}
{\bf ATLAS} Collaboration, {\it {Search for supersymmetry in events with large
  missing transverse momentum, jets, and at least one tau lepton in 20
  fb$^{-1}$ of $\sqrt{s}$=8 TeV proton-proton collision data with the ATLAS
  detector}},  \href{http://arxiv.org/abs/1407.0603}{{\tt arXiv:1407.0603}}.

\bibitem{atlasMMCitation28}
{\bf ATLAS} Collaboration, {\it {Measurement of the $t\bar{t}$ production
  cross-section as a function of jet multiplicity and jet transverse momentum
  in 7 TeV proton-proton collisions with the ATLAS detector}},
  \href{http://arxiv.org/abs/1407.0891}{{\tt arXiv:1407.0891}}.

\bibitem{cmsMMCitation0}
{\bf CMS} Collaboration, {\it {Measurement of the top-quark mass in $t\bar{t}$
  events with dilepton final states in $pp$ collisions at $\sqrt{s}=7$ TeV}},
  {\em Eur.Phys.J.} {\bf C72} (2012) 2202,
  [\href{http://arxiv.org/abs/1209.2393}{{\tt arXiv:1209.2393}}].

\bibitem{cmsMMCitation1}
{\bf CMS} Collaboration, {\it {Identification of b-quark jets with the CMS
  experiment}},  {\em JINST} {\bf 8} (2013) P04013,
  [\href{http://arxiv.org/abs/1211.4462}{{\tt arXiv:1211.4462}}].

\bibitem{cmsMMCitation2}
{\bf CMS} Collaboration, {\it {Search for a Higgs boson decaying into a b-quark
  pair and produced in association with b quarks in proton-proton collisions at
  7 TeV}},  {\em Phys.Lett.} {\bf B722} (2013) 207--232,
  [\href{http://arxiv.org/abs/1302.2892}{{\tt arXiv:1302.2892}}].

\bibitem{Chatrchyan:2012bra}
{\bf CMS} Collaboration, {\it {Measurement of the $t\bar{t}$ production cross
  section in the dilepton channel in $pp$ collisions at $\sqrt{s}=7$ TeV}},
  {\em JHEP} {\bf 1211} (2012) 067, [\href{http://arxiv.org/abs/1208.2671}{{\tt
  arXiv:1208.2671}}].

\bibitem{Mistlberger:2012rs}
B.~Mistlberger and F.~Dulat, {\it {Limit setting procedures and theoretical
  uncertainties in Higgs boson searches}},
  \href{http://arxiv.org/abs/1204.3851}{{\tt arXiv:1204.3851}}.

\bibitem{BIPM1995}
BIPM, IEC, IFCC, ISO, IUPAC, IUPAP, and OIML, {\em {Guide to the Expression of
  Uncertainty in Measurement}}.
\newblock International Organization for Standardization, 1995.

\bibitem{likelihoodMMCranmerComm}
{\bf ATLAS} Collaboration, K.~Cranmer. private communication, 2013.

\bibitem{James:1975dr}
F.~James and M.~Roos, {\it {Minuit: A System for Function Minimization and
  Analysis of the Parameter Errors and Correlations}},  {\em
  Comput.Phys.Commun.} {\bf 10} (1975) 343--367.

\bibitem{2011EPJC...71.1554C}
G.~{Cowan}, K.~{Cranmer}, E.~{Gross}, and O.~{Vitells}, {\it {Asymptotic
  formulae for likelihood-based tests of new physics}},  {\em European Physical
  Journal C} {\bf 71} (Feb., 2011) 1554,
  [\href{http://arxiv.org/abs/1007.1727}{{\tt arXiv:1007.1727}}].

\bibitem{2003sppp.conf..261C}
K.~{Cranmer}, {\it {Frequentist Hypothesis Testing with Background
  Uncertainty}},  in {\em Statistical Problems in Particle Physics,
  Astrophysics, and Cosmology} (L.~{Lyons}, R.~{Mount}, and R.~{Reitmeyer},
  eds.), p.~261, 2003.
\newblock \href{http://arxiv.org/abs/physics/0310108}{{\tt physics/0310108}}.

\bibitem{boxplots}
M.~Frigge, D.~C. Hoaglin, and B.~Iglewicz, {\it Some implementations of the
  boxplot},  {\em The American Statistician} {\bf 43} (1989), no.~1 pp. 50--54.

\end{thebibliography}\endgroup
\end{document}